\documentclass{kluwer}    

\newdisplay{guess}{Conjecture}

\begin{document}                                                                                   
\begin{article}
\begin{opening}         
\title{Migration of Trans-Neptunian Objects to the Terrestrial Planets} 
\author{Sergei I. \surname{Ipatov} }
\runningauthor{S.I. Ipatov and J.C. Mather}
\runningtitle{Migration of TNOs to the terrestrial planets}
\institute{George Mason University; NASA/GSFC; Institute of Applied Mathematics, Moscow}
\author{John C. \surname{Mather} }
\institute{NASA/GSFC}
\date{}

\begin{abstract}

The orbital evolution of more than 22000 Jupiter-crossing objects under the gravitational influence
of planets was investigated. We found that the mean collision probabilities of Jupiter-crossing objects 
(from initial orbits  close to the orbit of a comet) 
with the terrestrial planets can differ by more than two orders of magnitude for different comets.
For  initial orbital elements close to those of  some comets (e.g. 2P and 10P),  about
0.1\% of objects  got Earth-crossing orbits with
semi-major axes $a$$<$2 AU 
and moved in such orbits for more than a Myr (up to tens or even 
hundreds of Myrs). 
Results of our runs testify in favor of at least one of these conclusions:
1) the portion of 1-km former trans-Neptunian objects (TNOs) among near-Earth objects (NEOs)
 can exceed several tens of percents, 2) the number of TNOs migrating
inside solar system could be smaller by a factor of several than it was earlier considered,
3) most of 1-km former TNOs that had got NEO orbits disintegrated 
into mini-comets and dust during a smaller part of their dynamical lifetimes 
if these lifetimes are not small. 

\end{abstract}
\keywords{trans-Neptunian objects, Jupiter-family comets, terrestrial planets}

\end{opening}           

\section{Introduction}  
                  
Trans-Neptunian objects (TNOs) are considered to be one of
the main sources of near-Earth objects (NEOs). Bottke et al. 
(2002),
Binzel et al. (2002), and Weissman et al. (2002) 
believe that asteroids are the main source of NEOs.
Duncan et al. (1995) and Kuchner et al. (2002) 
investigated the migration of TNOs
to Neptune's orbit, and  Levison and Duncan (1997) studied their 
migration from Neptune's orbit to Jupiter's orbit. 
Based on the results of migration of 
Jupiter-crossing objects (JCOs) with initial orbits
close to the orbit of Comet P/1996 R2 obtained by Ipatov and Hahn (1999),
Ipatov (1999, 2001) found that 10-20\% or more of the 1-km 
Earth-crossers could have come from the Edgeworth-Kuiper belt into 
Jupiter-crossing orbits. 
In the present paper we 
consider a larger number of JCOs than before. Some preliminary results 
were presented by  Ipatov (2002a-b), 
who also discussed the 
formation of TNOs and asteroids. The results of the runs of JCOs,
including several figures, can be also found in (Ipatov and Mather, 2003a-b).
A wider review on the migration of asteroids and comets to NEO orbits
was made by Ipatov (2001). 

\section{Migration of Jupiter-Family Comets to the Terrestrial Planets}

As the migration of TNOs to Jupiter's orbit was investigated
by several authors, we have made a series of simulations of the orbital evolution 
of JCOs under the gravitational influence of 
planets. We omitted the influence of Mercury 
(except for Comet 2P/Encke) and Pluto.  The orbital evolution of more than 9000 and 
13000 JCOs with initial 
periods $P_a$$<$20 yr was integrated with the use of the Bulirsch-Stoer and 
symplectic methods (BULSTO and RMVS3 codes), respectively.
We used the integration package of Levison and Duncan (1994).  

      In the first series of runs (denoted as $n1$) we calculated the 
evolution of 3100 JCOs moving in initial orbits close to those of 20 
real comets with period $5$$<$$P_a$$<$9 yr, 
and in the second series of runs (denoted as $n2$) we considered 10000 JCOs 
moving in initial orbits close to those of 10 real comets (with numbers 
77, 81, 82, 88, 90, 94, 96, 97, 110, 113) with period 5$<$$P_a$$<$15 yr.
In other series of runs, initial 
orbits were close to those of a single comet (2P/Encke, 9P/Tempel 1, 10P/Tempel 2, 
22P/Kopff, 28P/Neujmin 1, 
39P/Oterma, or 44P/Reinmuth 2). 
In order to compare the orbital evolution of comets and asteroids,
we also investigated the orbital evolution of asteroids initially moving 
in the 3:1 and 5:2 resonances with Jupiter. For  JCOs we varied 
only the initial mean anomaly $\nu$ in an interval less than several tens of degrees.  
Usually in one run, there were 250 JCOs or 144 asteroids.
For asteroids, we varied initial values of $\nu$ and the 
longitude of the ascending node from 0 to 360$^\circ$.
The approximate values of initial semi-major axes, eccentricities
and inclinations of considered comets
($a_\circ$, $e_\circ$, and $i_\circ$) are presented in Table I.
We investigated the orbital evolution during the dynamical lifetimes of objects
(at least until all the objects reached perihelion distance $q$$>$6 AU).

\begin{table}
\caption[]{Semi-major axes (in AU), eccentricities and inclinations of several considered comets}
\begin{tabular}{lccc|lccc} 
 & $a_\circ$ & $e_\circ$ &$i_\circ$ & & $a_\circ$ & $e_\circ$& $i_\circ$ \\
\hline
2P/Encke & 2.22 & 0.85 & $11.7^\circ$ & 9P/Tempel 1 & 3.12 &0.52&$10.5^\circ$ \\
10P/Tempel 2 & 3.10 & 0.53 & $12.0^\circ$ &
 22P/Kopff & 3.47&0.54&$4.7^\circ$ \\
28P/Neujmin 1 & 6.91 & 0.78 & $14.2^\circ$ & 39P/Oterma & 7.25&0.25&$1.9^\circ$ \\
 44P/Reinmuth 2 & 3.53 & 0.46& 7.0$^\circ$  & 88P/Howell & 3.13 & 0.56 & 4.4$^\circ$ \\
96P/Machholz 1 & 3.04 & 0.96 & 60.2$^\circ$ & 113P/Spitaler & 3.69 & 0.42 & 5.8$^\circ$ 
\end{tabular} 
\end{table}

\begin{table*}
\begin{minipage}{12cm}

\caption[]{
Mean probability $P$$=$$10^{-6}P_r$ of a collision of an object with 
a planet (Venus=V, Earth=E, Mars=M)
during its lifetime,  mean time $T$ (in Kyr) during which 
$q$$<$$a_{pl}$, $T_c$=$T/P$ (in Gyr),
mean time $T_J$ (in Kyr) spent in Jupiter-crossing orbits, 
mean time $T_d$ (in Kyr) spent in orbits with $Q$$<$4.2 AU,
and ratio $r$ of times spent in Apollo and Amor orbits. 
Results from  BULSTO 
code at $10^{-9}$$\le$$\varepsilon$$\le$$10^{-8}$
(marked as $10^{-9}$) and 
at  $\varepsilon$$\le$$10^{-12}$ (marked as $10^{-12}$)
and with RMVS3 code (Levison and Duncan, 1994)
at integration step $d_s$.
In the case of asteroids, for the last four lines $e_\circ$=0.05 and $i_\circ$=$5^\circ$, and 
for other runs $e_\circ$=0.15 and $i_\circ$=$10^\circ$.

}

$ \begin{array}{lll|cccccccccl}
\hline

   & & &$V$ & $V$ &  $E$ & $E$ & $E$ & $M$ & $M$ &  & &\\

\cline{4-13}

 &\varepsilon $ or $ d_s & N& P_r & T & P_r & T &T_c& P_r & T & r & T_J &T_d \\ 

\hline

n1&10^{-9}&1900& 2.42 & 4.23 & 4.51 & 7.94 &1.76&  6.15 & 30.0 & 0.7 & 119& 20 \\ 
n1&\le$$10^d&1200&25.4 & 13.8 & 40.1 & 24.0 &0.60&2.48 & 35.2 
& 3.0 &117 & 25.7 \\ 
n1&\le$$10^d&1199&7.88 & 9.70 & 4.76 & 12.6 &2.65&0.76 & 16.8 
& 2.8 &117 &10.3 \\ 
\hline
n2&10^{-9}&1000&10.2&27.5&14.7&43.4&2.95&2.58&62.6&3.1&187&8.3\\
n2&\le$$10^d&9000&15.3&25.4&15.0&37.0&2.47&2.75&57.3&3.1&148&19.9\\
\hline
$2P$&10^{-9} & 501 & 141 &345 & 110 & 397 &3.61&10.5& 430 & 18.& 173&249 \\

$2P$&10^{-12}&100 &321&541&146&609&4.2   &14.8 & 634 & 27. &20 & 247\\ 
$2P$&10^d & 251 &860&570&  2800 &788 & 0.28 &294 & 825 & 22. &0.29&614 \\ 
$2P$&10^d & 250 &160&297&  94.2 & 313 & 3.32 & 10.0&324 &35. &0.29& 585\\ 
\hline
$9P$&10^{-9} & 800&1.34 &1.76 &3.72 & 4.11 &1.10&0.71 & 9.73& 1.2 &96& 2.6 \\ 
$9P$&10^d & 400 &1.37&3.46&3.26&7.84&2.40&1.62&23.8& 1.1 &128& 8.0 \\ 
\hline
$10P$&10^{-9} & 2149&28.3 & 41.3& 35.6 & 71.0&1.99&10.3 & 169.&1.6 &122&107\\ 

$10P$&\le$$10^d &450&14.9&30.4&22.4&41.3&1.84&6.42& 113.&1.5 &85&44.\\ 
\hline
$22P$& 10^{-9} &1000&1.44&2.98&1.76&4.87&2.77&0.74 & 11.0& 1.6 &116&1.5 \\ 
$22P$&10^d & 250& 0.68 &2.87&1.39&4.96&3.57&0.60 & 11.5& 1.5 &121 & 0.6 \\ 
\hline
$28P$&10^{-9} & 750 & 1.7 & 21.8& 1.9 & 34.7&18.3& 
0.44 & 68.9&1.9 &443&0.1 \\ 
$28P$&10^d & 250 & 3.87 & 35.3& 3.99 & 59.0 &14.8& 0.71 & 109. 
&2.2 &535 &3.3 \\ 
\hline
$39P$&10^{-9} & 750 & 1.06&1.72&1.19& 3.03&2.55& 
0.31 & 6.82& 1.6 &94&2.7 \\ 
$39P$&10^d & 250 & 2.30 &2.68&2.50& 4.22 & 1.69 & 0.45 & 7.34& 
 2.2 &92 & 0.5 \\ 
\hline
$44P$&10^{-9} & 500 &2.58&15.8&4.01&24.9&6.21&0.75&46.3&2.0&149&8.6\\
$44P$&10^d&1000&3.91&5.88&5.84&9.69&1.66&0.77&16.8&2.3&121&2.9\\
\hline

\hline
3:1&10^{-9}& 288  &1286 &1886& 1889& 2747 &1.45& 
488&4173 & 2.7 &229&5167 \\
3:1&10^{-12}& 70 & 1162 &1943& 1511& 5901 &3.91& 587& 
803& 4.6 &326 & 8400 \\
3:1&10^d& 142 &27700&8617& 2725& 9177& 
3.37& 1136&9939& 16. &1244&5000 \\
\hline
5:2&10^{-9}& 288 &101   &173& 318 & 371 &1.16&209 
&1455 & 0.5 &233& 1634 \\
5:2&10^{-12}& 50 & 130   &113& 168 & 230 &1.37&46.2 &507 
& 1.4 &166 &512 \\
5:2&10^d& 144 & 58.6&86.8&86.7&174 &2.01&17.&355&1.7 &224& 828\\

\hline
3:1&10^{-9}& 144 & 200 &420&417 &759  &1.82& 195&1423&2.1 &157&2620 \\
3:1&10^d& 144 & 10051&2382&6164&4198&0.68&435&5954&2.5&235&18047\\
5:2&10^{-9}& 144 & 105 &114& 146& 214 &1.47& 42&501& 1.5 &193&996 \\
5:2&10^d& 144 &148&494&173&712&4.12&51&1195&2.3&446&984\\
\hline

\end{array} $

\end{minipage}

\end{table*}

      In our runs, planets were considered as material points, so 
literal collisions did not occur.  However, 
based on the orbital elements sampled with a 500 yr 
step, we calculated the mean probability  $P$ of collisions. We 
define $P$ as  $P_\Sigma/N$, where $P_\Sigma$ is the total probability
of collisions of $N$ objects  with a planet during their lifetimes,   the mean 
time $T$=$T_\Sigma/N$ during which
perihelion distance $q$ of an object was less than the semi-major 
axis $a_{pl}$ of the planet,
the mean time $T_d$  spent in orbits with aphelion distance $Q$$<$4.2 AU,
and the mean time $T_J$ during which an object moved in 
Jupiter-crossing orbits. The values of $P_r$=$10^6P$,
$T_J$, $T_d$, and $T$ are shown in Table II. Here $r$ is the ratio of the 
total time interval when orbits are of Apollo
type ($a$$>$1 AU, $q$$<$$1.017$ AU) at $e$$<$0.999 to that 
of Amor type ($1.017$$<$$q$$<$1.3 AU) and $T_c$=$T/P$ (in Gyr).
In almost all runs $T$ was equal to the mean time 
in orbits which cross the orbit of the planet and 1/$T_c$ was a probability
of a collision per year. 

In Table II we present the results obtained by
the Bulirsch-Stoer method  with the integration step 
error less than
$\varepsilon$$\in$[$10^{-9}$-$10^{-8}$] and also with 
$\varepsilon$$\le$$10^{-12}$ and  by a symplectic method 
with an integration step $d_s$$\le$10 days. 
For these three series of runs, the  results obtained were similar
(except for  probabilities of close encounters with the Sun
when they were high). 
For $d_s$=30 days for most of the objects we found similar results,
but we found a larger portion of the objects that reached Earth-crossing orbits
with $a$$<$2 AU for several tens of Myr and even inner-Earth orbits (IEOs, i.e.
with $Q$$<$0.983 AU).
These few bodies increased the mean values of $P$ by a factor of
more than 10, and the mean probabilities were greater than for $d_s$$\le$10 days.

The results can differ considerably depending on the initial orbits of comets. 
The values of $P$ for Earth were about (1-4)$\times$$10^{-6}$ for Comets 9P,
22P, 28P, and 39P. For Comet 10P they were greater by an order of magnitude
than for 9P, though initial orbits of 9P and 10P were close. 
This is a real difference in dynamics of two comets and is not "luck of the draw"
in the integrations.
$P$ exceeded $10^{-4}$ for Comet 2P.

The probability of a collision with Earth (or with Venus and Mars) for one object
that orbited for several Myr with  $Q$$<$4.2 AU
could be much greater than the total probability for hundreds other objects. 
Some had typical asteroidal and NEO orbits and reached $Q$$<$3 AU for several Myr. 
One object with initial orbit close to that of Comet 88P/Howell 
after 40 Myr got $Q$$<$3.5 AU and moved in
orbits with $a$$\approx$2.60-2.61 AU, 1.7$<$$q$$<$2.2 AU, 3.1$<$$Q$$<$3.5 AU,
$e$$\approx$0.2-0.3, and $i$$\approx$5-10$^\circ$ for 650 Myr. 
If we consider this object, then 
for series $n2$ at $d_s$$\le$10$^d$ the value of $T_d$ will 
be greater by a factor of 4 (i.e., $\approx$80 Kyr)  than that in 
the corresponding line  of Table II.
The times spent by five specific objects that have large probabilities of collisions
with the terrestrial planets while in IEO, Aten, 
Al2 (1$<$$a$$<$2 AU, $q$$<$1.017 AU), Apollo, and Amor orbits
are presented in Table III.

With RMVS3 at $d_s$$\le$10 days for 2P run, the value of $P$ for Earth for one object
presented in  line 1 of Table III  was greater
by a factor of 30  than for 250 other objects (see Table II).
For series $n1$ with RMVS3, the
probability of a collision with Earth for 
one object with initial orbit close to that of Comet 44P/Reinmuth 2 was 88.3\%
of the total probability for 1200 objects from this series, and the 
total probability for 1198 objects was only 4\%.
This object (line 2 in Table III) was not included in Table II with $N$=1199 for $n1$. 

\begin{table}[h]

\begin{center}
\begin{minipage}{12cm}
\caption{Times (in Myr) spent by five objects in various 
orbits and probabilities of their collisions
 with Venus ($p_v$), Earth ($p_e$), and Mars ($p_m$)
during their lifetimes $T_{lt}$ (in Myr). }
$ \begin{array}{ll| lll lllll l}
\hline

$Comet$ & d_s $ or $ \varepsilon&$IEOs$& $Aten$&  $Al2$& $Apollo$ & $Amor$ 
& T_{lt} &p_v&p_e&p_m\\
\hline
$       2P$ & 10^d & 12 & 33.6 & 73.4 & 75.6 & 4.7 & 126 &0.18&0.68&0.07\\
$       44P$ & 10^d&  0 & 0    & 11.7 & 14.2 & 4.2 & 19.5 &0.02&0.04&0.002\\
$       2P$ & 10^{-8}& 0.1& 83 &249 & 251 &15  & 352 &0.224&0.172&0.065\\
$       10P$ & 10^{-8}& 10& 3.45 &0.06 &0.06  &0.05  & 13.6 &0.655&0.344&0.001\\
$      113P$ &6^d&0 &0 & 56.8& 59.8&4.8& 67& 0.037& 0.016&0.0001\\

\hline

\end{array}$
\end{minipage}
\end{center}
\end{table}

For BULSTO at $\varepsilon$$\in$[$10^{-9}$-$10^{-8}$] two objects
(lines 3-4 in Table III)
with the largest probabilities were not included in Table II
for 2P at $N$=501, and for 10P at $N$=2149.
The probabilities of collisions of these two objects with Earth and Venus
(see Table III) were greater than for 9350 other objects combined (0.17 for Earth and 0.15
for Venus).
Large values of  $P$ for Mars in the $n1$ runs with BULSTO were caused by a single object with a 
lifetime of 26 Myr. 
Ipatov (1995) obtained the migration of JCOs into IEO and Aten 
orbits using the approximate
method of spheres of action for taking into account the gravitational 
interactions of bodies with planets.


The times spent by  22000 JCOs in Earth-crossing orbits 
with $a$$<$2 AU were  due to 
a few tens of objects with high collision probabilities.
With BULSTO at $10^{-9}$$\le$$\varepsilon$$\le$$10^{-8}$
 six and nine objects, respectively from 10P and 2P series, 
moved into Apollo orbits with 
$a$$<$2 AU (Al2 orbits) for at least 0.5 Myr each, and five of them 
remained in such orbits for more than 5 Myr each. The contribution of 
all the 9337 other objects to Al2 orbits was smaller. 
Among the 9352 JCOs considered with BULSTO,
only one and two JCOs reached IEO and Aten orbits, respectively.
Only one object in series $n2$ (line 5 in Table III) got Al2 orbits 
during more than 1 Myr.

For the $n1$ series of runs, while moving in JCO orbits,
objects had orbital periods $P_a$$<$20 yr (Jupiter-family comets)
and 20$<$$P_a$$<$200 yr (Halley-type comets)
for 32\% and 38\% of $T_J$=0.12 Myr, respectively. 

Some former JCOs spent a long time in the 3:1 resonance with Jupiter
and with 2$<$$a$$<$2.6 AU.
Other objects reached Mars-crossing orbits 
for long times. So JCOs can supply bodies to the 
regions which are considered by many scientists (Bottke et al., 2002) 
to belong to the main sources of NEOs.
The probabilities of collisions of bodies with the Earth per unit of time, i.e.
the values of 1/$T_c$, were of the same order for JCOs and resonant asteroids.
Therefore, mean eccentricities and inclinations of Earth-crossers were 
similar for former TNOs and resonant asteroids.
With BULSTO the mean probability of collisions with the Earth for the 5:2 resonance 
 was  1/3 of that for the 3:1 resonance at $e_\circ$=0.05 and 
this difference was greater by a factor of several at $e_\circ$=0.15
(see Table II). 

The ratio $P_S$ of the number of objects colliding with the Sun to 
the total number of
escaped (collided or ejected) objects was less than 0.015 
for the simulations, except for Comet 2P/Encke, Comet 96P/Machholz 1 from $n2$ series, 
and resonant asteroids. In the case
of close encounters with the Sun, the values of $P_S$ obtained by BULSTO and
RMVS3 and at different $\varepsilon$ and $d_s$ were different, but all 
other results were similar, as probabilities of collisions
of objects with the terrestrial planets were usually small
after their close encounters with the Sun.

The results presented in the paper were obtained
for direct modelling of collisions with the Sun, 
but usually they are practically the same
if we consider that objects disappear when perihelion distance $q$ becomes
less than the radius $r_S$ of the Sun or even several such radii 
(i.e., we checked $q$$<$$k_S r_S$, where $k_S$ equals 0, 1, or another value).
The only noticeable difference was for Comet 96P from $n2$ series
and a smaller one for Comet 2P.
For $n2$ series, several runs, in which there was an appreciable 
difference in time spent in orbits with $Q$$<$4.7 AU for $k_S$=0
and for $k_S$=1 (the times can differ by a factor of several), 
were not included in Tables 2 and 4. This difference
was due to Comet 96P. Eccentricity and inclination of this comet 
are large, and they become even larger after close encounters with the Sun,
so usually even for these runs the collision probabilities of 
objects with the terrestrial planets were not differed much (by more than
15\%) at $k_S$=0 and $k_S$=1. 
There were
three runs, 
for each of which
at $k_S$=0 a body in orbit close to that of Comet 96P was responsible for
70-75\% of collision probabilities  with the Earth, 
and for $k_S$=1 a lifetime of such body
was much less than for $k_S$=0. Nevertheless, for all ($\sim 10^4$) objects
from $n2$ series, at $k_S$=0  the probabilities 
of collisions with the terrestrial planets were close to those at $k_S$=1,
even if we consider the above runs. The difference for times spent
in Earth-crossing orbits is greater than that for the probabilities and
is about 20\%.
For all runs at 2P series, 
the difference in time spent in orbits with $Q$$<$4.7 AU for $k_S$=0
and for $k_S$=1 was less than 4\%.
In 2P series
of runs (and also for the 3:1 resonance with Jupiter), 
at $k_S$=0 we sometimes got orbits with $i$$>$$90^\circ$, but practically 
there were no such orbits at $k_S$$\ge$1 (Ipatov and Mather, 2003a-b).
For Comet 96P we found one object which also got $i$$>$90$^\circ$
for 3 Myr. Inclinations of other orbits initially close to the orbit 
of this comet did not exceed 90$^\circ$.

\section{Trans-Neptunian Objects in Near-Earth Object Orbits}

Using the results of migration of TNOs obtained by Duncan et al. (1995),
considering the total of $5\times10^9$ 1-km 
TNOs with $30$$<$$a$$<$$50$ AU (Jewit and Fernandez, 2001), and assuming
that the mean time for a body to move in a Jupiter-crossing orbit is about
0.12 Myr, Ipatov (2001) found that about $N_{Jo}$=$10^4$ 1-km former TNOs are now 
Jupiter-crossers, and 3000 are Jupiter-family comets.
Using the total times spent by $N$
simulated JCOs in various orbits, we obtained the following 
numbers of 1-km former TNOs now moving in several types of orbits:  

\begin{table}[h]
\begin{center}
\begin{minipage}{11cm}
\caption{Estimates of the number of 1-km former TNOs now moving in several ty\-pes of orbits}

$ \begin{array}{ccc| ccccc}
\hline
$$N$$ &$method$& $series$ & $IEOs$ & $Aten$ & $Al2$ & $Apollo$ & $Amor$ \\
\hline
$3100$ & $BULSTO, RMVS3$ & $n1$  & 0   & 0   & 480 & 1250 & 900 \\
$8000$ & $RMVS3$ & $n2$ & 0 & 0 & 500 & 2800 & 800 \\
$8800$ &$BULSTO$ & $without 2P$& 95 &  30 & 230 & 2600 & 1560\\
$9352$ &$BULSTO$ & $all$& 90 & 770 & 3700 & 6500 & 1700\\
\hline

\end{array} $
\end{minipage}
\end{center}
\end{table}

For example, the number of IEOs $N_{IEOs}$=$N_{Jo}t_{IEO}$/$(N_J t_J)$,
where $t_{IEO}$ is the total time during which $N_J$ former JCOs 
moved in IEO orbits, and $N_J t_J$ is the total time during which 
$N_J$ JCOs moved in Jupiter-crossing orbits.
The number of former TNOs in Apollo and Amor orbits can be estimated on 
the basis of $n1$ and  $n2$ runs. 
The number of NEOs with diameter $d$$\ge$1 km is estimated to be
about 1500 (Rabinowitz et al., 1994) or 1000 (Morbidelli et al., 2002). 
 Half of NEOs are Earth-crossers.
Even if the number of Apollo objects is 
smaller by a factor of several than that based on $n1$ and $n2$ runs,
it is comparable to the real number (500-750) of 1-km
Earth-crossing objects (half of them are in orbits with $a$$<$2 AU), 
although the
latter number does not include those in highly eccentric orbits.
The portions of objects in Aten and Al2 orbits
are much greater in our 2P runs than in other runs. 
Our estimates of these portions are very approximate.
The above estimates of the portion of former TNOs in NEO orbits are
relatively large (up to tens of percents), but it is also possible 
that the number of TNOs migrating inside solar 
system could be smaller by a factor of several than it was earlier considered.

Comets are estimated to be active for $T_{act}$$\sim$$10^3$--$10^4$ yr. 
$T_{act}$ is smaller for closer encounters with the Sun (Weissman et al., 2002),
so for Comet 2P it is smaller than for other Jupiter-family comets.
Some former comets can move for tens or
even hundreds of Myr in NEO and asteroidal orbits,  so the number of extinct comets 
can exceed the number of active comets by several orders of magnitude.
The mean time spent by Encke-type objects in Earth-crossing orbits is 
$\ge$0.4 Myr.
This time corresponds to $\ge$40-400  
extinct comets of this type. Note that the diameter of Comet 2P 
is about 5-8 km (Fernandez et al., 2000; Lowry et al., 2003), 
so the number of 1-km Earth-crossing extinct comets can 
exceed 1000.
The rate of a cometary object decoupling from the Jupiter vicinity 
and transferring to an NEO-like orbit can be increased by a
factor of several due to nongravitational effects
(Asher et al., 2001; Fernandez and Gallardo, 2002). 

      Based on the collision probability $P=4\times10^{-6}$ 
we find that  1-km former TNOs  collide with the Earth once in 3 Myr. 
This value of $P$ is smaller
than that for our $n1$ and $n2$ runs and does not include the 
`champions' in collision probability.  Using $P=4\times10^{-6}$ 
and assuming that the total mass of planetesimals that ever crossed Jupiter's 
orbit is $\sim$$ 100m_\oplus$, where
$m_\oplus$ is the mass of the Earth (Ipatov, 1993), 
we concluded
that the total mass of water delivered from the feeding zone of
the giant planets to the Earth could be about the mass of Earth oceans.
  

Our runs showed that if one observes
former comets in NEO orbits, then  most of them could have already
moved in such orbits for millions 
of years.
Some former comets that have moved in typical NEO orbits for millions 
or even hundreds of millions of years, and might have had
multiple close encounters with the Sun,
could have lost their mantles,
which caused their low albedo, and so change their albedo (for most
observed NEOs, the albedo is greater than that for comets; Fernandez et al., 2001)
and would look like typical asteroids, or some of them could disintegrate into
mini-comets and dust.

Chen and Jewitt (1994) noted that
while cometary splitting is sometimes associated with the close
passage of a comet by the Sun, it is also known to occur at
heliocentric distances of up to 9 AU. 
At 10m, the near-Earth flux is more than two orders of magnitude 
greater than power law extrapolated from larger sizes.
Levison et al. (2002) obtained that majority of comets evolved
inward from the Oort cloud must physically disrupt, but
Jupiter-family comets do not appear to disrupt at the same rate. 
Bailey (2002) consider that some long-period comets
become innert 
and hence evolve into low-albedo objects resembling asteroids,
and another alternative is that Oort cloud comets may easily break up 
into 
unobserved smaller bodies or dust.

%

From measured albedos, Fernandez et al. (2001) concluded that
the fraction of extinct comets among NEOs and unusual asteroids is significant
($\ge$9\%).
Rickman et al. (2001) and Jewitt and Fernandez (2001) considered that 
dark spectral classes that might include the 
ex-comets are severely under-represented and comets played an important 
and perhaps even dominant role among all km-size 
Earth impactors.


  {\bf We conclude} that the trans-Neptunian belt can 
provide a significant portion of the Earth-crossing objects, 
or the number of TNOs migrating inside solar 
system could be smaller than it was earlier considered,
or most of 1-km former TNOs that had got NEO orbits disintegrated into 
mini-comets and dust during a smaller part of their dynamical lifetimes if these 
lifetimes are not small. 


      This work was supported by NRC (0158730), NASA (NAG5-10776), 
INTAS (00-240), and RFBR (01-02-17540). We thank the anonymous referee
for helpful remarks. 

\end{article}
\end{document}